\newcommand{\un}{~\mathrm}
\newcommand{\ie}{{\em i.e. }}
\newcommand{\eg}{{\em e.g. }}

\documentclass[prl
,twocolumn
,floatfix
,preprintnumbers
,showpacs
,amsmath
,amssymb
,superscriptaddress]{revtex4}
\usepackage[T1]{fontenc}
\usepackage{graphicx}
\usepackage{dcolumn}
\usepackage{bm}
\usepackage[dvips]{color}

\begin{document}
\title{Brittle/quasi-brittle transition in dynamic fracture: An energetic signature}

\author{J. Scheibert}
\altaffiliation{Present address: PGP, University of Oslo, Oslo, Norway}
\email{julien.scheibert@fys.uio.no}
\affiliation{CEA, IRAMIS, SPCSI, Grp. Complex Systems $\&$ Fracture, F-91191 Gif sur Yvette, France}
\affiliation{Unité Mixte CNRS/Saint-Gobain, Surface du Verre et Interfaces, 39 Quai Lucien Lefranc, 93303 Aubervilliers cedex, France}
\author{C. Guerra}
\affiliation{CEA, IRAMIS, SPCSI, Grp. Complex Systems $\&$ Fracture, F-91191 Gif sur Yvette, France}
\affiliation{Facultad de Ingeniería Mecánica y Eléctrica, Universidad Autónoma de Nuevo León, Ave. Universidad, S/N, Ciudad Universitaria, C.P. 66450, San Nicolás de los Garza, NL, Mexico}
\author{F. Célarié}
\altaffiliation{Present address: LARMAUR, Univ. of Rennes 1, France}
\affiliation{CEA, IRAMIS, SPCSI, Grp. Complex Systems $\&$ Fracture, F-91191 Gif sur Yvette, France}
\affiliation{Unité Mixte CNRS/Saint-Gobain, Surface du Verre et Interfaces, 39 Quai Lucien Lefranc, 93303 Aubervilliers cedex, France}
\author{D. Dalmas}
\affiliation{Unité Mixte CNRS/Saint-Gobain, Surface du Verre et Interfaces, 39 Quai Lucien Lefranc, 93303 Aubervilliers cedex, France}
\author{D. Bonamy}
\affiliation{CEA, IRAMIS, SPCSI, Grp. Complex Systems $\&$ Fracture, F-91191 Gif sur Yvette, France}

\pacs{46.50.+a, 
62.20.M-, 
61.43.-j 
}

\begin{abstract}
Dynamic fracture experiments were performed in PMMA over a wide range of velocities and reveal that the fracture energy exhibits an abrupt 3-folds increase from its value at crack initiation at a well-defined critical velocity, below the one associated to the onset of micro-branching instability. This transition is associated with the appearance of conics patterns on fracture surfaces that, in many materials, are the signature of damage spreading through the nucleation and growth of micro-cracks. A simple model allows to relate both the energetic and fractographic measurements. These results suggest that dynamic fracture at low velocities in amorphous materials is controlled by the brittle/quasi-brittle transition studied here.
\end{abstract}

\maketitle

Dynamic fracture drives catastrophic material failures. Over the last century, a coherent theoretical framework, the so-called Linear Elastic Fracture Mechanics (LEFM) has developed and provides a quantitative description of the motion of a single smooth crack in a linear elastic material \cite{Freund-CUP-1990}. LEFM assumes that all the mechanical energy released during fracturing is dissipated at the crack tip. Defining the fracture energy $\Gamma$ as the energy needed to create two crack surfaces of a unit area, the instantaneous crack growth velocity $v$ is then selected by the balance between the energy flux and the dissipation rate $\Gamma v$. This yields \cite{Freund-CUP-1990}:
\begin{eqnarray}
\Gamma \simeq \left( 1-v/c_R \right) {K^2(c)}/{E}, \label{Eq1}
\end{eqnarray}
\noindent where $c_R$ and $E$ are the Rayleigh wave speed and the Young modulus of the material, respectively, and $K(c)$ is the Stress Intensity Factor (SIF) for a quasi-static crack of length $c$. $K$ depends only on the applied loading and specimen geometry, and characterizes entirely the stress field in the vicinity of the crack front.

Equation (\ref{Eq1}) describes quantitatively the experimental results for dynamic brittle fracture at slow crack velocities \cite{Bergkvist-EngFractMech-1974}. However, large discrepancies are observed in brittle amorphous materials at high velocities \cite{Fineberg-Marder-PhysRep-1999, RaviChandar-Elsevier-2004, Livne-BenDavid-Fineberg-PhysRevLett-2007, Livne-Bouchbinder-Fineberg-PhysRevLett-2008}. In particular (i) the measured maximum crack speeds lie in the range $0.5-0.6c_R$, i.e. far smaller than the limiting speed $c_R$ predicted by Eq. (\ref{Eq1}) and (ii) fracture surfaces become rough at high velocities (see \cite{Fineberg-Marder-PhysRep-1999, RaviChandar-Elsevier-2004} for reviews). It has been argued \cite{Sharon-Fineberg-Nature-1999} that experiments start to depart from theory above a critical $v_b \simeq 0.4 c_R$ associated to the onset of micro-branching instabilities \cite{Fineberg-Gross-Marder-Swinney-PhysRevLett-1991}: for $v > v_b$ the crack motion becomes a multi-cracks state. This translates into (i) a dramatic increase of the fracture energy $\Gamma$ at $v_b$, due to the increasing number of micro-branches propagating simultaneously and (ii) a non-univocal relation between $\Gamma$ and $v$ \cite{Sharon-Fineberg-Nature-1999}. The micro-branching instability hence yielded many recent theoretical efforts \cite{branching}.
However, a number of puzzling observations remain at smaller velocities. In particular, even for velocities much lower than $v_b$, (i) the measured dynamic fracture energy is generally much higher than that at crack initiation \cite{Sharon-Fineberg-Nature-1999, Kalthoff-Winkler-Beinert-IntJFract-1976, Rosakis-Duffy-Freund-JMechPhysSolids-1984, Bertram-Kalthoff-Materialprofung-2003} and (ii) fracture surfaces roughen over length scales much larger than the microstructure scale ("mist" patterns) \cite{Hull-CUP-1999}, the origin of which remains debated \cite{mist, Lawn-CUP-1993}.

In this Letter, we report dynamic fracture experiments in polymethylmethacrylate (PMMA), the archetype of brittle amorphous materials, designed to unravel the primary cause of these last discrepancies. We show that dynamic fracture energy exhibits an abrupt 3-folds increase from its value at crack initiation at a well-defined critical velocity $v_a$ well below $v_b$. This increase coincides with the onset of damage spreading through the nucleation and growth of micro-cracks, the signature of which is the presence of conic patterns on post-mortem fracture surfaces. A simple model for this nominally brittle to quasi-brittle transition is shown to reproduce both the energetic and fractographic measurements.

Dynamic cracks are driven in PMMA with measured Young modulus and Poisson ratio of $E=2.8\pm 0.2 \un{GPa}$ and $\nu = 0.36$, which yields $c_R = 880 \pm 30\un{m.s^{-1}}$. Its fracture energy at the onset of crack propagation was determined to be $K_c^2/E=0.42 \pm 0.07 \un{kJ.m^{-2}}$, with $K_c$ being the material toughness. Specimen are prepared from $140 \times 125 \times 15\un{mm}^3$ parallelepipeds in the $x$ (propagation), $y$ (loading) and $z$ (thickness) directions by cutting a $25 \times 25\un{mm}^2$ rectangle from the middle of one of the $125 \times 15\un{mm}^2$ edges and then cutting a $10\un{mm}$ groove deeper into the specimen (Fig. 1, bottom inset). Two steel jaws equipped with rollers are placed on both sides of the cut-out rectangle and a steel wedge (semi-angle $15^\circ$) is pushed between them at constant velocity $38~\mu\mathrm{m.s^{-1}}$ up to crack initiation. In this so-called wedge splitting geometry, the SIF $K$ decreases with the crack length $c$. To increase its value at crack initiation, and therefore the initial crack velocity, a circular hole with a radius ranging between $2$ and $8\un{mm}$ is drilled at the tip of the groove to tune the stored mechanical energy $U_0$. Dynamic crack growth with instantaneous velocities ranging from $75\un{m.s^{-1}}$ to {\bf $500\un{m.s^{-1}}$} and stable trajectories are obtained. The location $c(t)$ of the crack front is measured during each experiment ($40~\mu\mathrm{m}$ and $0.1~\mu\mathrm{s}$ resolutions) using a modified version of the potential drop technique: A series of 90 parallel conductive lines ($2.4\un{nm}$-thick Cr layer covered with $23\un{nm}$-thick Au layer), $500\un\mu\mathrm{m}$-wide with an $x$-period of $1\un{mm}$ are deposited on one of the $x$-$y$ surfaces of the specimen, connected in parallel and alimented with a voltage source. As the crack propagates, the conductive lines are cut at successive times, these events being detected with an oscilloscope. The instantaneous crack velocity $v(c)$ is computed from $c(t)$, and the instantaneous SIF $K(c)$ is calculated using 2D finite element calculations (software Castem 2007) on the exact experimental geometry, assuming plane stress conditions and a constant wedge position as boundary condition.

\begin{figure}
\begin{center}
\includegraphics[width=0.95\columnwidth]{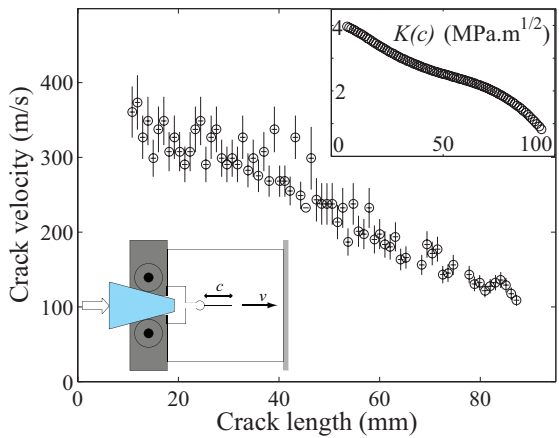}
\caption{Measured crack velocity $v$ as a function of crack length $c$ in a typical experiment ($U_0 = 2.6\un{J}$). The vertical lines are error bars. Top inset: Calculated quasi-static SIF $K$ as a function of $c$. Bottom inset: Schematics of the Wedge-Splitting test.\label{fig1}}
\end{center}
\end{figure}

\begin{figure}
\begin{center}
\includegraphics[width=0.95\columnwidth]{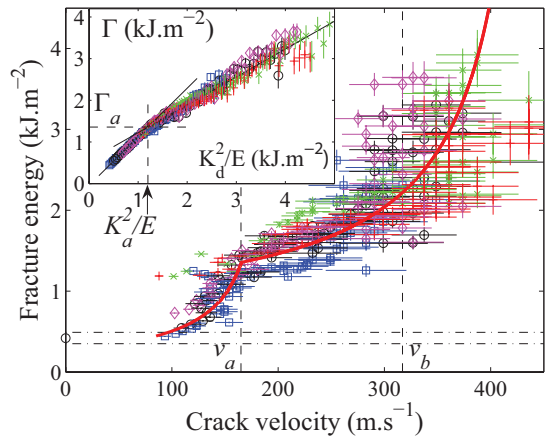}
\caption{(color online). Fracture energy $\Gamma$ as a function of crack velocity $v$ for five different experiments with different stored mechanical energies $U_0$ at crack initiation: 2.0 ($\square$), 2.6 ($\circ$), 2.9 ($\lozenge$), 3.8 ($+$) and 4.2$\un{J}$ ($\times$). The two vertical dashed lines correspond to $v_a$ and $v_b$. The two horizontal dashed lines indicate the confidence interval for the measured fracture energy $K_c^2/E$ at crack initiation. Thick red line: model prediction. Inset: $\Gamma$ as a function of $K_d^2/E$ (see model) for the same experiments. A crossover between two linear regimes (linear fits in black lines) occurs at ($K_d^2/E = K_a^2/E\simeq 1.2\un{kJ.m^{-2}}$ ; $\Gamma=\Gamma_a\simeq 1.34\un{kJ.m^{-2}}$).\label{fig2}}
\end{center}
\end{figure}

Values for the fracture energy $\Gamma$ are obtained directly from Eq. (\ref{Eq1}) by combining the $v$ measurements and the $K$ calculations. Typical $v(c)$ and $K(c)$ curves are shown in Fig. \ref{fig1}. The variations of $\Gamma$ with $v$ (Fig. \ref{fig2}) are found to be the same in various experiments performed with various stored mechanical energy $U_0>2.0 \un J$ at crack initiation. This curve provides evidence for three regimes, separated by two critical velocities. For slow crack velocities, $\Gamma$ remains of the order of $K_c^2/E$ as expected in LEFM. Then, as $v$ reaches the first critical velocity $v_a \simeq 165 m.s^{-1} = 0.19 c_R$, $\Gamma$ increases abruptly to a value about 3 times larger than $K_c^2/E$. Beyond $v_a$, $\Gamma$ increases slowly with $v$ up to the second critical velocity, $v_{b} = 0.36 c_R \simeq 317\un{m.s^{-1}}$ \cite{Sharon-Fineberg-Nature-1999}, above which $\Gamma$ diverges again with $v$. This second increase corresponds to the onset of the micro-branching instability, widely discussed in the literature \cite{Fineberg-Gross-Marder-Swinney-PhysRevLett-1991, Sharon-Fineberg-Nature-1999}, whereas the first one, at $v_a$, is reported here for the first time. The high slope of $\Gamma(v)$ around $v_a$ provides a direct interpretation for the repeated observation of cracks that span a large range of $\Gamma$ but propagate at a nearly constant velocity of about $0.2 c_R$ (see \eg refs. \cite{RaviChandar-Knauss-IntJFract-1984-3, RaviChandar-Yang-JMechPhysSolids-1997}).

\begin{figure}[!ht]
\begin{center}
\includegraphics[width=0.95\columnwidth]{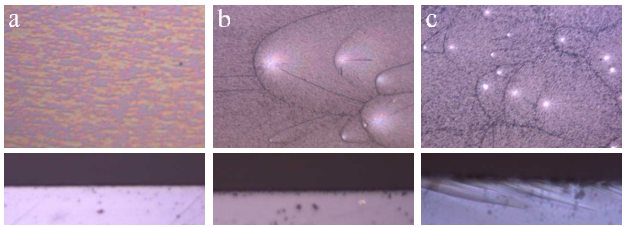}
\caption{Microscope images ($\times 10$) taken at (a) $v=120\pm20\un{m.s^{-1}}$, $K^2/E=1\un{kJ.m^{-2}}$ (b) $v=260\pm30\un{m.s^{-1}}$, $K^2/E=2\un{kJ.m^{-2}}$ (c) $v=650 \pm 100\un{m.s^{-1}}$ ($K^2/E=7\un{kJ.m^{-2}}$). Top line : fracture surfaces ($0.5 \times 0.7 \un{mm}^2$ field of view). Bottom line : sample sides ($0.25 \times 0.7 \un{mm}^2$ field of view). Crack propagation is from left to right. \label{fig3}}
\end{center}
\end{figure}

The post-mortem fracture surfaces shed light on the nature of the transition at $v=v_a$ on the curve $\Gamma(v)$. Fig. \ref{fig3} shows the surface morphology for increasing crack velocity. For $v<v_a$, the fracture surfaces remain smooth at the optical scale (Fig. \ref{fig3}(a), top). Above $v_a$ conic marks are observed (Figs. \ref{fig3}(b) and \ref{fig3}(c), top). They do not leave any visible print on the sides of the specimens (Fig. \ref{fig3}(b), bottom), contrary to the micro-branches that develop for $v \geq v_b$ (Fig. \ref{fig3}(c), bottom).

Similar conic marks were reported in the fracture of many other amorphous brittle materials (see \cite{RaviChandar-Elsevier-2004, Hull-CUP-1999} and references therein), including polymer glasses, silica glasses and polycrystals. Their formation is thought to arise from inherent toughness fluctuations at the micro-structure scale due to material heterogeneities randomly distributed in the material \cite{Smekal-OsterrIngArch-1953, RaviChandar-Yang-JMechPhysSolids-1997}. The enhanced stress field in the vicinity of the main crack front activates some of the low toughness zones and triggers the initiation of secondary penny-shaped micro-cracks ahead of the crack front. Each micro-crack grows radially under the stress associated with the main crack along a plane different from it. When two cracks intersect in space and time, the ligament separating them breaks up, leaving a visible conic marking on the post-mortem fracture surface.

\begin{figure}[!ht]
\begin{center}
\includegraphics[width=0.95\columnwidth]{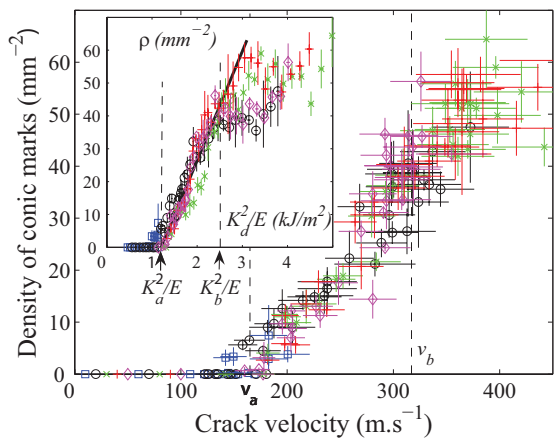}
\caption{(color online). Surface density $\rho$ of conic marks as a function of crack velocity for all experiments shown in Fig. \ref{fig2}. Inset: $\rho$ as a function of $K_d^2/E$ (linear fit in black line).\label{fig4}} 
\end{center}
\end{figure}

Figure \ref{fig4} shows the surface density of conic marks $\rho$ as a function of crack velocity $v$. Below $v_a$, no conic mark is observed up to $\times 50$ magnification, consistently with \cite{Sheng-Zhao-IntJFract-1999}. Above $v_a$, $\rho$ increases almost linearly with $v-v_a$. The exact correspondence between the critical velocity $v_a$ at which $\Gamma$ exhibits an abrupt increase and the velocity at which the first conic marks appear on the fracture surfaces strongly suggests that both phenomena are associated with the same transition. The nucleation and growth of micro-cracks can therefore be identified as the new fracture mechanism that starts at $v_a$. This damage process is generic in brittle materials and is relevant for an even wider range of materials than those that exhibit conic marks, \eg granite \cite{Moore-Lockner-JStructGeol-1995}.

We now present a simple model reproducing the $\Gamma (v)$ curve between 0 and $v_b$. We assume that linear elasticity fails in the material when the local stress reaches a yield stress $\sigma_Y$. It defines a fracture process zone (FPZ) around the crack tip, the size of which is given by $R_c(v)=K_d^2(c,v)/a\sigma_Y^2$ where $a$ is a dimensionless constant \cite{Lawn-CUP-1993} and $K_d$ is the dynamic SIF. We consider that all the dissipative phenomena (plastic deformations, crazing or cavitation for instance) occur in the FPZ, with a volumic dissipated energy $\epsilon$. The material is then assumed to contain a volume density $\rho_s$ of discrete "source-sinks" (SS, see \eg \cite{Lawn-CUP-1993} for previous uses of this concept). Each SS is assumed to activate into a micro-crack if two conditions are met: (i) the local stress reaches $\sigma_Y$ and (ii) the SS is located at a distance from the crack tip larger than $d_a$ \cite{note1}. The nucleation of a micro-crack is assumed to be accompanied by an excluded volume $V$ where stress is screened \ie no SS can acivate anymore. In the following, $\rho_s$, $\sigma_Y$, $\epsilon$, $d_a$ and $V$ are taken as constants throughout the material. Three cases should be considered:

\noindent (I) - At the onset of crack propagation,  all the volume within $R_c(v=0)=K_c^2/ a\sigma_Y^2$ contributes to the fracture energy $\Gamma(v=0)=K_c^2/E$.

\noindent (II) - For $v \leq v_a$, no micro-crack nucleates and $R_c(v)={K_d(c,v)}^2/ a\sigma_Y^2<d_a$. The dynamic SIF is then $K_d(c,v)=k(v)K(c)$ \cite{Freund-CUP-1990} where $k(v)\simeq(1-v/c_R)/\sqrt{1-v/c_D}$ is universal and $c_D$ is the dilatational wave speed (here $c_D$=2010$\pm$60$\un{m.s^{-1}}$). The volume scanned by the FPZ when the crack surface increases by $S$ is $2 R_c(v) S$. The dissipated energy $\Gamma(v) S$ is given by $\gamma S+ 2 \epsilon R_c(v) S$ where $\gamma$ is the Griffith surface energy. Since $\Gamma(v=0)=K_c^2/E$, one finally gets for $v \leq v_a$:
\begin{align}
\Gamma(v)=\alpha\frac{K_d(v)^2}{E}+\left(1-\alpha\right) \frac{K_c^2}{E} \quad \mathrm{with} \quad \alpha=\frac{2 \epsilon E}{a\sigma_Y^2}. \label{Eq2}
\end{align}
This predicted linear dependence of $\Gamma$ with ${K_d}^2/E$ for $v \leq v_a$ is in agreement with measurements (Fig. \ref{fig2}, inset). A linear fit to the data (correlation coefficient $R$=0.985) gives $\alpha$=1.17$\pm$0.05 and $K_c^2/E$=0.3$\pm$0.2$\un{kJ/m^{2}}$, where $\pm$ stands for 95$\%$ confidence interval. The latter value is compatible with the measurements of the fracture energy at crack initiation. By combining Eqs. (\ref{Eq1}) and (\ref{Eq2}), one gets a prediction for the $\Gamma (v)$ curve \cite{unpublished} that reproduces very well the low velocity regime in Fig. \ref{fig2}. Extrapolation of this regime \cite{unpublished} exhibits a divergence of the dissipated energy for a finite velocity $v_a'$=$(\alpha-1)c_R c_D/(\alpha c_D - c_R)\simeq$200$\un{m.s^{-1}}\simeq$0.23$c_R$, slightly larger than $v_a$. In the absence of micro-cracks, this velocity $v_a'$ would have therefore set the limiting macroscopic crack velocity.

\noindent (III) - For $v \geq v_a$, $R_c(v)\geq d_a$ \ie micro-cracks start to nucleate. The surface density of micro-cracks $\rho(v)$ is then equal to the number of activated SS beyond $d_a$ per unit of fracture area, \ie $\rho_s \left\{2\left[R_c(v)-R_c(v_a)\right]-\rho V\right\}$ where the third term in the parenthesis stands for the excluded sites around micro-cracks. This yields:
\begin{align}
\rho(v) = \beta \frac{K_d(v)^2-K_a^2}{E} \quad \mathrm{with} \quad \beta=\frac{2 E}{a \sigma_Y^2} \frac{\rho_s}{1+\rho_s V}\label{Eqrho}
\end{align}
where $K_a$=$K_d(v_a)$. This linear relationship is in good agreement with the measurements for $\rho(K_d^2/E)$ before the micro-branching onset, beyond which $\rho$ saturates (Fig. \ref{fig4}, inset). A fit to the data ($R$=0.877) between $K_a$ and $K_b=K_d(v_b)$ gives $\beta$ =33$\pm$3$\un{J^{-1}}$. In the micro-cracking regime, the local dynamic SIF $K_d$ is not equal to the macroscopic one anymore, but corresponds to that at the individual micro-crack tips, at which the limiting velocity is expected to be $v_a' \gtrsim v_a$. It is then natural to assume that all micro-cracks propagate at the same velocity $v_a$, which yields $K_d(v)=k(v_a)K$ \cite{note2}. The energy $\Gamma(v) S$ dissipated when the crack surface increases by $S$ is $\gamma S+\epsilon \left[2 R_c(v) S-\rho(v) S V\right]$, yielding:
\begin{align}
\Gamma(v) = \Gamma_a+ \chi \frac{K_d(v)^2-K_a^2}{E} \quad \mathrm{with} \quad \chi=\frac{{2 \epsilon E}/{a \sigma_Y^2}}{1+\rho_s V} \label{Eq3}
\end{align}
where $\Gamma_a=\Gamma(v_a)$. Eq. (\ref{Eq3}) predicts a linear dependence of $\Gamma$ with $K_d^2/E$, in agreement with the measurements for $K_d^2/E>K_a^2/E$ (Fig. \ref{fig2}, inset). A linear fit to the data between $K_a$ and $K_b=K_d(v_b)$ gives $\chi=0.67\pm0.01$. The corresponding predicted $\Gamma (v)$ curve \cite{unpublished} reproduces very well the intermediate velocity regime $v_a<v<v_b$ (Fig. \ref{fig2}) and exhibits a divergence of the dissipated energy for $v_\infty = c_R(1-\chi k(v_a)^2) \simeq 450\un{m.s}^{-1}\simeq 0.52 c_R$. This limiting velocity is very close to the observed maximum crack speed in brittle amorphous materials.

This simple scenario allows to illustrate how material defects control the dynamic fracture of amorphous solids before the onset of micro-branching. For $v<v_a$, the mechanical energy released at the crack tip is dissipated into both a constant surfacic energy and a volumic energy within the FPZ, the size of which increases with crack speed. With this mechanism alone, the crack speed would be limited to a value slightly larger than $v_a$. But damage spreading through micro-cracking makes possible to observe much larger velocities: The crack propagates through the nucleation, growth and coalescence of micro-cracks, with a macroscopic effective velocity that can be much larger than the local velocity of each micro-crack tip \cite{RaviChandar-Yang-JMechPhysSolids-1997, Prades-Bonamy-Dalmas-Bouchaud-Guillot-IntJSolidsStruct-2005}. We suggest that micro-cracks in themselves do not increase dissipation, but rather decrease it by locally screening the stress. At velocities larger than $v_b$, micro-branches contribute to the dissipated energy proportionally to their surface \cite{Sharon-Gross-Fineberg-PhysRevLett-1996}. We emphasize that the {\em nominally brittle to quasi-brittle transition} occurring at $v_a$ is very likely to be generic for amorphous solids and should therefore be taken into account in future conceptual and mathematical descriptions of dynamic fracture. In this respect, Continuum Damage Mechanics (CDM) \cite{Kachanov-Springer-1986} initially derived for "real" quasi-brittle materials like ceramics or concrete may be relevant to describe fast crack growth in nominally brittle materials. In particular, a better understanding of the relationship between the dynamics of propagation of both the individual micro-cracks and the macroscopic crack is still needed.

We thank P. Viel and M. Laurent (SPCSI) for gold deposits, T. Bernard (SPCSI) for technical support, K. Ravi-Chandar (University of Texas, Austin) and J. Fineberg (The Hebrew University of Jerusalem) for fruitful discussions and P. Meakin (INL/PGP) for a careful reading of the manuscript. We acknowledge funding from French ANR through Grant No. ANR-05-JCJC-0088 and from Mexican CONACYT through Grant No. 190091.


\begin{thebibliography}{31}
\expandafter\ifx\csname natexlab\endcsname\relax\def\natexlab#1{#1}\fi
\expandafter\ifx\csname bibnamefont\endcsname\relax
  \def\bibnamefont#1{#1}\fi
\expandafter\ifx\csname bibfnamefont\endcsname\relax
  \def\bibfnamefont#1{#1}\fi
\expandafter\ifx\csname citenamefont\endcsname\relax
  \def\citenamefont#1{#1}\fi
\expandafter\ifx\csname url\endcsname\relax
  \def\url#1{\texttt{#1}}\fi
\expandafter\ifx\csname urlprefix\endcsname\relax\def\urlprefix{URL }\fi
\providecommand{\bibinfo}[2]{#2}
\providecommand{\eprint}[2][]{\url{#2}}

\bibitem[{\citenamefont{Freund}(1990)}]{Freund-CUP-1990}
\bibinfo{author}{\bibfnamefont{L.}~\bibnamefont{Freund}},
  \emph{\bibinfo{title}{Dynamic Fracture Mechanics}}
  (\bibinfo{publisher}{Cambridge University Press, Cambridge, England}, \bibinfo{year}{1990}).

\bibitem[{\citenamefont{Bergkvist}(1974)}]{Bergkvist-EngFractMech-1974}
\bibinfo{author}{\bibfnamefont{H.}~\bibnamefont{Bergkvist}},
  \bibinfo{journal}{Eng. Fract. Mech.} \textbf{\bibinfo{volume}{6}},
  \bibinfo{pages}{621} (\bibinfo{year}{1974}).

\bibitem[{\citenamefont{Fineberg and
  Marder}(1999)}]{Fineberg-Marder-PhysRep-1999}
\bibinfo{author}{\bibfnamefont{J.}~\bibnamefont{Fineberg}} \bibnamefont{and}
  \bibinfo{author}{\bibfnamefont{M.}~\bibnamefont{Marder}},
  \bibinfo{journal}{Phys. Rep.} \textbf{\bibinfo{volume}{313}},
  \bibinfo{pages}{1} (\bibinfo{year}{1999}).

\bibitem[{\citenamefont{Ravi-Chandar}(2004)}]{RaviChandar-Elsevier-2004}
\bibinfo{author}{\bibfnamefont{K.}~\bibnamefont{Ravi-Chandar}},
  \emph{\bibinfo{title}{Dynamic Fracture}} (\bibinfo{publisher}{Elsevier, Amsterdam},
  \bibinfo{year}{2004}).

\bibitem[{\citenamefont{Livne et~al.}(2007)\citenamefont{Livne, Ben-David, and
  Fineberg}}]{Livne-BenDavid-Fineberg-PhysRevLett-2007}
\bibinfo{author}{\bibfnamefont{A.}~\bibnamefont{Livne}},
  \bibinfo{author}{\bibfnamefont{O.}~\bibnamefont{Ben-David}},
  \bibnamefont{and} \bibinfo{author}{\bibfnamefont{J.}~\bibnamefont{Fineberg}},
  \bibinfo{journal}{Phys. Rev. Lett.} \textbf{\bibinfo{volume}{98}},
  \bibinfo{pages}{124301} (\bibinfo{year}{2007}).

\bibitem[{\citenamefont{Livne et~al.}(2008)\citenamefont{Livne, Bouchbinder,
  and Fineberg}}]{Livne-Bouchbinder-Fineberg-PhysRevLett-2008}
\bibinfo{author}{\bibfnamefont{A.}~\bibnamefont{Livne}},
  \bibinfo{author}{\bibfnamefont{E.}~\bibnamefont{Bouchbinder}},
  \bibnamefont{and} \bibinfo{author}{\bibfnamefont{J.}~\bibnamefont{Fineberg}},
  \bibinfo{journal}{Phys. Rev. Lett.} \textbf{\bibinfo{volume}{101}},
  \bibinfo{pages}{264301} (\bibinfo{year}{2008}).

\bibitem[{\citenamefont{Sharon and
  Fineberg}(1999)}]{Sharon-Fineberg-Nature-1999}
\bibinfo{author}{\bibfnamefont{E.}~\bibnamefont{Sharon}} \bibnamefont{and}
  \bibinfo{author}{\bibfnamefont{J.}~\bibnamefont{Fineberg}},
  \bibinfo{journal}{Nature} \textbf{\bibinfo{volume}{397}},
  \bibinfo{pages}{333} (\bibinfo{year}{1999}).

\bibitem[{\citenamefont{Fineberg et~al.}(1991)\citenamefont{Fineberg, Gross,
  Marder, and Swinney}}]{Fineberg-Gross-Marder-Swinney-PhysRevLett-1991}
\bibinfo{author}{\bibfnamefont{J.}~\bibnamefont{Fineberg}},
\bibinfo{author}{et al.}
  \bibinfo{journal}{Phys. Rev. Lett.} \textbf{\bibinfo{volume}{67}},
  \bibinfo{pages}{457} (\bibinfo{year}{1991}).

\bibitem{branching}
\bibinfo{author}{\bibfnamefont{M.}~\bibnamefont{Adda-Bedia}},
  \bibinfo{journal}{Phys. Rev. Lett.} \textbf{\bibinfo{volume}{93}},
  \bibinfo{pages}{185502} (\bibinfo{year}{2004});
\bibinfo{author}{\bibfnamefont{H.}~\bibnamefont{Henry}} \bibnamefont{and}
  \bibinfo{author}{\bibfnamefont{H.}~\bibnamefont{Levine}},
  \bibinfo{journal}{Phys. Rev. Lett.} \textbf{\bibinfo{volume}{93}},
  \bibinfo{pages}{105504} (\bibinfo{year}{2004});
\bibinfo{author}{\bibfnamefont{E.}~\bibnamefont{Bouchbinder}},
  \bibinfo{author}{\bibfnamefont{J.}~\bibnamefont{Mathiesen}},
  \bibnamefont{and}
  \bibinfo{author}{\bibfnamefont{I.}~\bibnamefont{Procaccia}},
  \bibinfo{journal}{Phys. Rev. E} \textbf{\bibinfo{volume}{71}},
  \bibinfo{pages}{056118} (\bibinfo{year}{2005});
\bibinfo{author}{\bibfnamefont{H.}~\bibnamefont{Henry}}, \bibinfo{journal}{EPL}
  \textbf{\bibinfo{volume}{83}}, \bibinfo{pages}{16004} (\bibinfo{year}{2008}).

\bibitem[{\citenamefont{Kalthoff et~al.}(1976)\citenamefont{Kalthoff, Winkler,
  and Beinert}}]{Kalthoff-Winkler-Beinert-IntJFract-1976}
\bibinfo{author}{\bibfnamefont{J.~F.} \bibnamefont{Kalthoff}},
  \bibinfo{author}{\bibfnamefont{S.}~\bibnamefont{Winkler}}, \bibnamefont{and}
  \bibinfo{author}{\bibfnamefont{J.}~\bibnamefont{Beinert}},
  \bibinfo{journal}{Int. J. Fract.} \textbf{\bibinfo{volume}{12}},
  \bibinfo{pages}{317} (\bibinfo{year}{1976}).

\bibitem[{\citenamefont{Rosakis et~al.}(1984)\citenamefont{Rosakis, Duffy, and
  Freund}}]{Rosakis-Duffy-Freund-JMechPhysSolids-1984}
\bibinfo{author}{\bibfnamefont{A.~J.} \bibnamefont{Rosakis}},
  \bibinfo{author}{\bibfnamefont{J.}~\bibnamefont{Duffy}}, \bibnamefont{and}
  \bibinfo{author}{\bibfnamefont{L.~B.} \bibnamefont{Freund}},
  \bibinfo{journal}{J. Mech. Phys. Solids} \textbf{\bibinfo{volume}{32}},
  \bibinfo{pages}{443} (\bibinfo{year}{1984}).

\bibitem[{\citenamefont{Bertram and
  Kalthoff}(2003)}]{Bertram-Kalthoff-Materialprofung-2003}
\bibinfo{author}{\bibfnamefont{A.}~\bibnamefont{Bertram}} \bibnamefont{and}
  \bibinfo{author}{\bibfnamefont{J.~F.} \bibnamefont{Kalthoff}},
  \bibinfo{journal}{Materialprüfung} \textbf{\bibinfo{volume}{45}},
  \bibinfo{pages}{100} (\bibinfo{year}{2003}).

\bibitem[{\citenamefont{Hull}(1999)}]{Hull-CUP-1999}
\bibinfo{author}{\bibfnamefont{D.}~\bibnamefont{Hull}},
  \emph{\bibinfo{title}{Fractography}} (\bibinfo{publisher}{Cambridge University Press, Cambridge, England},
  \bibinfo{year}{1999}).

\bibitem{mist}
\bibinfo{author}{\bibfnamefont{J.~W.}~\bibnamefont{Johnson}},\bibnamefont{and}
  \bibinfo{author}{\bibfnamefont{D.~G.}~\bibnamefont{Holloway}},
  \bibinfo{journal}{Phil. Mag.} \textbf{\bibinfo{volume}{14}},
  \bibinfo{pages}{731} (\bibinfo{year}{1966});
\bibinfo{author}{\bibfnamefont{T.}~\bibnamefont{Cramer}},
  \bibinfo{author}{\bibfnamefont{A.}~\bibnamefont{Wanner}}, \bibnamefont{and}
  \bibinfo{author}{\bibfnamefont{P.}~\bibnamefont{Gumbsch}},
  \bibinfo{journal}{Phys. Rev. Lett.} \textbf{\bibinfo{volume}{85}},
  \bibinfo{pages}{788} (\bibinfo{year}{2000});
\bibinfo{author}{\bibfnamefont{D.}~\bibnamefont{Bonamy}} \bibnamefont{and}
  \bibinfo{author}{\bibfnamefont{K.}~\bibnamefont{Ravi-Chandar}},
  \bibinfo{journal}{Phys. Rev. Lett.} \textbf{\bibinfo{volume}{91}},
  \bibinfo{pages}{235502} (\bibinfo{year}{2003});
\bibinfo{author}{\bibfnamefont{M.~J.} \bibnamefont{Buehler}} \bibnamefont{and}
  \bibinfo{author}{\bibfnamefont{H.}~\bibnamefont{Gao}},
  \bibinfo{journal}{Nature} \textbf{\bibinfo{volume}{439}},
  \bibinfo{pages}{307} (\bibinfo{year}{2006});
\bibinfo{author}{\bibfnamefont{G.}~\bibnamefont{Wang}},
\bibinfo{author}{et al.}
  \bibinfo{journal}{Phys. Rev. Lett.} \textbf{\bibinfo{volume}{98}},
  \bibinfo{pages}{235501} (\bibinfo{year}{2007});
\bibinfo{author}{\bibfnamefont{A.}~\bibnamefont{Rabinovitch}} \bibnamefont{and}
  \bibinfo{author}{\bibfnamefont{D.}~\bibnamefont{Bahat}},
  \bibinfo{journal}{Phys. Rev. E} \textbf{\bibinfo{volume}{78}},
  \bibinfo{pages}{067102} (\bibinfo{year}{2008}).

\bibitem[{\citenamefont{Lawn}(1993)}]{Lawn-CUP-1993}
\bibinfo{author}{\bibfnamefont{B.}~\bibnamefont{Lawn}},
  \emph{\bibinfo{title}{Fracture of Brittle Solids}}
  (\bibinfo{publisher}{Cambridge University Press, Cambridge, England}, \bibinfo{year}{1993}).

\bibitem[{\citenamefont{Ravi-Chandar and
  Knauss}(1984)}]{RaviChandar-Knauss-IntJFract-1984-3}
\bibinfo{author}{\bibfnamefont{K.}~\bibnamefont{Ravi-Chandar}}
  \bibnamefont{and} \bibinfo{author}{\bibfnamefont{W.~G.}
  \bibnamefont{Knauss}}, \bibinfo{journal}{Int. J. Fract.}
  \textbf{\bibinfo{volume}{26}}, \bibinfo{pages}{141} (\bibinfo{year}{1984}).

\bibitem[{\citenamefont{Ravi-Chandar and
  Yang}(1997)}]{RaviChandar-Yang-JMechPhysSolids-1997}
\bibinfo{author}{\bibfnamefont{K.}~\bibnamefont{Ravi-Chandar}}
  \bibnamefont{and} \bibinfo{author}{\bibfnamefont{B.}~\bibnamefont{Yang}},
  \bibinfo{journal}{J. Mech. Phys. Solids} \textbf{\bibinfo{volume}{45}},
  \bibinfo{pages}{535 } (\bibinfo{year}{1997}).

\bibitem[{\citenamefont{Smekal}(1953)}]{Smekal-OsterrIngArch-1953}
\bibinfo{author}{\bibfnamefont{A.}~\bibnamefont{Smekal}},
  \bibinfo{journal}{Oesterr. Ing. Arch.} \textbf{\bibinfo{volume}{7}},
  \bibinfo{pages}{49} (\bibinfo{year}{1953}).

\bibitem[{\citenamefont{Sheng and Zhao}(1999)}]{Sheng-Zhao-IntJFract-1999}
\bibinfo{author}{\bibfnamefont{J.~S.} \bibnamefont{Sheng}} \bibnamefont{and}
  \bibinfo{author}{\bibfnamefont{Y.~P.} \bibnamefont{Zhao}},
  \bibinfo{journal}{Int. J. Fract.} \textbf{\bibinfo{volume}{98}},
  \bibinfo{pages}{L9} (\bibinfo{year}{1999}).

\bibitem[{\citenamefont{Moore and
  Lockner}(1995)}]{Moore-Lockner-JStructGeol-1995}
\bibinfo{author}{\bibfnamefont{D.~E.} \bibnamefont{Moore}} \bibnamefont{and}
  \bibinfo{author}{\bibfnamefont{D.~A.} \bibnamefont{Lockner}},
  \bibinfo{journal}{J. Struct. Geol.} \textbf{\bibinfo{volume}{17}},
  \bibinfo{pages}{95} (\bibinfo{year}{1995}).

\bibitem{note1}
We believe that conic marks correspond to the fraction of micro-cracks having sufficient time to develop up to optical scale. For too small FPZ (for $v \leq v_a$), nucleated micro-cracks are rapidly caught up by the main crack, only leaving undetectable submicrometric elliptic marks.

\bibitem[{\citenamefont{unpublished}()}]{unpublished}
\bibinfo{author}{\bibfnamefont{C.}~\bibnamefont{Guerra}},
\bibinfo{author}{et al.,}{\bibnamefont{to be published}}.

\bibitem{note2}
This assumption was previously made (\cite{RaviChandar-Yang-JMechPhysSolids-1997} and references therein) and is fully consistent with the observed shape of conic marks in our experiments \cite{unpublished}.

\bibitem[{\citenamefont{Prades et~al.}(2005)\citenamefont{Prades, Bonamy,
  Dalmas, Bouchaud, and
  Guillot}}]{Prades-Bonamy-Dalmas-Bouchaud-Guillot-IntJSolidsStruct-2005}
\bibinfo{author}{\bibfnamefont{S.}~\bibnamefont{Prades}},
\bibinfo{author}{et al.}
  \bibinfo{journal}{Int. J. Solids Struct.} \textbf{\bibinfo{volume}{42}},
  \bibinfo{pages}{637} (\bibinfo{year}{2005}).

\bibitem[{\citenamefont{Sharon et~al.}(1996)\citenamefont{Sharon, Gross, and
  Fineberg}}]{Sharon-Gross-Fineberg-PhysRevLett-1996}
\bibinfo{author}{\bibfnamefont{E.}~\bibnamefont{Sharon}},
  \bibinfo{author}{\bibfnamefont{S.~P.} \bibnamefont{Gross}}, \bibnamefont{and}
  \bibinfo{author}{\bibfnamefont{J.}~\bibnamefont{Fineberg}},
  \bibinfo{journal}{Phys. Rev. Lett.} \textbf{\bibinfo{volume}{76}},
  \bibinfo{pages}{2117} (\bibinfo{year}{1996}).

\bibitem[{\citenamefont{Kachanov}(1986)}]{Kachanov-Springer-1986}
\bibinfo{author}{\bibfnamefont{L.~M.} \bibnamefont{Kachanov}},
  \emph{\bibinfo{title}{Introduction to Continuum Damage Mechanics}}
  (\bibinfo{publisher}{Martinus Nijhoff Publishers, Dordrecht}, \bibinfo{year}{1986}).

\end{thebibliography}
\end{document}